\documentclass[
               fleqn,usenatbib,useAMS]{mnras}

\usepackage{txfonts}
\usepackage[T1]{fontenc}
\usepackage{ae,aecompl}

\usepackage{graphicx}
\usepackage{savesym}
\savesymbol{iint}
\savesymbol{iiint}
\savesymbol{iiiint}
\savesymbol{idotsint}
\usepackage{amsmath}
\usepackage{amssymb}
\usepackage{gensymb}
\usepackage{bm}
\usepackage{aas_macros}
\usepackage[]{units}
\usepackage{multirow}
\usepackage{url}
\usepackage{color}
\usepackage[flushleft]{threeparttable}
\usepackage{tabularx}

\newcommand{\herschel}{\emph{Herschel}-SPIRE~} %
\newcommand{\pdv}{{\smallint}p\mathrm{d}V}

\newcommand{\pdvval}{(2.2 \pm 0.6)}
\newcommand{\pdvtotval}{(6.2 \pm 1.7)}
\newcommand{\pdvchisq}{36.3}
\newcommand{\pdvpte}{0.28}
\newcommand{\fval}{(14.5 \pm 3.3)}
\newcommand{\fchisq}{29.2}
\newcommand{\fpte}{0.61}

\newcommand{\nullchisq}{51.0}
\newcommand{\pdvnulldeltachisq}{14.7}
\newcommand{\fnulldeltachisq}{21.8}

\title[Evidence for SZ in Quasars]{Evidence for the Thermal Sunyaev-Zel'dovich Effect Associated
  with Quasar Feedback}

\author[D. Crichton et al.]{Devin Crichton$^{1}$\thanks{E-mail: dcrichton@jhu.edu},
Megan B. Gralla$^{1,2}$,
Kirsten Hall$^{1}$,
Tobias A. Marriage$^{1}$,
\newauthor
 Nadia L. Zakamska$^{1}$,
 Nick Battaglia$^{3}$,
J. Richard Bond$^{4}$,
Mark J. Devlin$^{5}$,
J. Colin Hill$^{6}$,
\newauthor
Matt Hilton$^{7}$,
Adam D. Hincks$^{8}$,
Kevin M. Huffenberger$^{9}$,
John P. Hughes$^{10}$,
\newauthor
Arthur Kosowsky$^{11}$,
Kavilan Moodley$^{7}$,
Michael D. Niemack$^{12}$,
Lyman A. Page$^{13}$,
\newauthor
Bruce Partridge$^{14}$,
Jonathan L. Sievers$^{15}$,
Crist\'obal Sif\'on$^{16}$,
Suzanne T. Staggs$^{13}$,
\newauthor
Marco P. Viero$^{17}$
and
Edward J. Wollack$^{18}$
\\
$^{1}$
Department of Physics and Astronomy,
Johns Hopkins University,
Baltimore, MD 21218, USA\\
$^{2}$
Harvard-Smithsonian Center for Astrophysics,
60 Garden Street,
Cambridge, MA 02138, USA\\
$^{3}$
Department of Astrophysical Sciences, Princeton University, Princeton,
NJ 08544, USA\\
$^{4}$
Canadian Institute for Theoretical Astrophysics,
University of Toronto,
Toronto, ON, M5S 3H8, Canada\\
$^{5}$
Department of Physics and Astronomy,
University of Pennsylvania,
209 South 33rd Street, Philadelphia, PA 19104 USA\\
$^{6}$
Department of Astronomy,
Pupin Hall,
Columbia University,
New York, NY 10027, USA\\
$^{7}$
Astrophysics and Cosmology Research Unit,
School of Mathematics, Statistics and Computer Science,\\
University of KwaZulu-Natal,
Durban 4041, South Africa\\
$^{8}$
Department of Physics and Astronomy,
University of British Columbia,
Vancouver, BC, V6T 1Z4, Canada\\
$^{9}$
Department of Physics,
Florida State University,
PO Box 3064350, Tallahassee, FL 32306-4350, USA\\
$^{10}$
Department of Physics and Astronomy,
Rutgers University,
136 Frelinghuysen Road,
Piscataway, NJ 08854-8019, USA\\
$^{11}$
 Department of Physics and Astronomy,
University of Pittsburgh,
Pittsburgh, PA, 15260, USA\\
$^{12}$
Department of Physics,
Cornell University,
Ithaca, NY 14853, USA\\
$^{13}$
Joseph Henry Laboratories of Physics, Jadwin Hall,
Princeton University,
Princeton, NJ, 08544, USA\\
$^{14}$
Department of Physics and Astronomy,
Haverford College,
Haverford, PA, 19041, USA\\
$^{15}$
Astrophysics and Cosmology Research Unit, School of Chemistry and Physics,
University of KwaZulu-Natal,
Durban 4041, South Africa\\
$^{16}$
Leiden Observatory,
Leiden University,
PO Box 9513, NL-2300 RA Leiden, Netherlands\\
$^{17}$
Kavli Institute for Particle Astrophysics and Cosmology,
Stanford University,
382 Via Pueblo Mall, Stanford, CA 94305, USA\\
$^{18}$
NASA/Goddard Space Flight Center,
Greenbelt, MD, 20771, USA
}

\date{Accepted 2016 February 10. Received 2016 January 29; in original form 2015 October 20}

\pubyear{2016}

\begin{document}
\label{firstpage}
\pagerange{\pageref{firstpage}--\pageref{lastpage}}
\maketitle

\begin{abstract}
  Using a radio-quiet subsample of the Sloan Digital Sky Survey spectroscopic quasar catalogue,
  spanning redshifts 0.5--3.5, we derive the mean millimetre and far-infrared quasar spectral energy
  distributions (SEDs) via a stacking analysis of Atacama Cosmology Telescope and
  \emph{Herschel}-Spectral and Photometric Imaging REceiver data. We constrain the form of the
  far-infrared emission and find 3$\sigma$--4$\sigma$ evidence for the thermal Sunyaev-Zel'dovich
  (SZ) effect, characteristic of a hot ionized gas component with thermal energy
  $\pdvtotval~\times~10^{60}$~erg. This amount of thermal energy is greater than expected assuming
  only hot gas in virial equilibrium with the dark matter haloes of
  $(1-5)\times~10^{12}h^{-1}$M$_\odot$ that these systems are expected to occupy, though the highest
  quasar mass estimates found in the literature could explain a large fraction of this energy. Our
  measurements are consistent with quasars depositing up to $\fval~\tau_8^{-1}$ per cent of their
  radiative energy into their circumgalactic environment if their typical period of quasar activity
  is $\tau_8\times~10^8$~yr. For high quasar host masses, $\sim10^{13}h^{-1}$M$_\odot$, this
  percentage will be reduced. Furthermore, the uncertainty on this percentage is only statistical
  and additional systematic uncertainties enter at the 40 per cent level. The SEDs are dust
  dominated in all bands and we consider various models for dust emission. While sufficiently
  complex dust models can obviate the SZ effect, the SZ interpretation remains favoured at the
  3$\sigma$--4$\sigma$ level for most models.
\end{abstract}

\section{Introduction}
\label{sec:intro}

Feedback from accreting black holes has become a key element in modelling galaxy evolution
\citep{tabo93, silk98, spri05}. Quasar feedback is routinely invoked in galaxy formation models to
quench star formation and to explain the steep decline of the bright end of the luminosity function
of galaxies \citep{thou95, crot06} and to reheat the intracluster medium \citep[e.g.,][]{rawl04,
  scan04}. The discovery that the masses of supermassive black holes in inactive galaxies strongly
correlate with the velocity dispersions and masses of their hosts' stellar bulges \citep{mago98,
  gebh00, ferr00} also suggests that the energy output of the black hole in its quasar phase must be
somehow coupled to the gas from which the stars form \citep{hopk06}. Finally, the similarity of the
history of star formation activity and that of supermassive black hole accretion in the Universe
\citep{boyl98, hopk08} suggests that these two phenomena physically affect each other or result from
the same physical process. Both star formation activity and black hole accretion activity peak at
$z>2$ and rapidly decline afterward; thus it is at these high redshifts that the physical connection
between supermassive black holes and their hosts was established.

How can supermassive black holes so profoundly affect their large-scale galactic and even
intergalactic environments? Even a small fraction of the binding energy of material accreted by the
supermassive black hole is in principle sufficient to liberate the galaxy-scale gas from the galaxy
potential, but coupling this energy output to the surrounding gas is no trivial task. However, there
are circumstances in which accretion energy clearly has had an impact on its environment. Radio jets
in massive elliptical galaxies and brightest cluster galaxies deposit energy into the hot gas
envelope \citep{mcna07, fabi12}. Likewise, powerful radio jets entrain warm gas and carry
significant amounts of material out of their host galaxies, especially at high redshifts
\citep{vanb86, tadh91, vill99, nesv06, nesv08, fu09}. While a narrow collimated outflow might seem
inefficient at impacting and removing the surrounding matter, in practice the power of the jet is
quickly and efficiently thermalized if it propagates into high density material \citep{bege89}.

Radio-quiet quasars (those that do not seem to have powerful radio jets) can also have a strong
effect on their large-scale environment. Radiation pressure, especially through absorption in
bound-bound transitions \citep{murr95, prog00}, can accelerate material in the immediate vicinity of
the black hole ($\la 1000$ Schwarzschild radii) to velocities $\la 10\%$ of the speed of light. This
material then impacts the interstellar medium in the host galaxy \citep{fauc12b, zubo12, nims14},
drives shocks into the surrounding medium, accelerates and destroys high-density clouds
\citep{macl94}, and over the lifetime of the quasar the resulting wind can engulf the entire
galaxy. High-velocity outflows on nuclear scales are observed in a large fraction of all quasars
\citep{weym81, turn84, cren03, reic03, gall07}, but evidence for large-scale impact of these
outflows has long been elusive. Only the last few years have brought about a flood of
observations of powerful, galaxy-wide outflows launched by radio-quiet quasars \citep{arav08, moe09,
alex10, gree11, cano12, liu13a, liu13b, hain13, rupk13a, veil13, harr14, hain14, zaka14,
brus15a, carn15, harr15, pern15, brus15b}.

Quasar-driven winds are likely to be inhomogeneous, with different phases of the wind medium
observable in different domains of the electromagnetic spectrum. The simplest models for these winds
involve hot, low density, volume-filling plasma \citep{fauc12b, zubo12}, with higher density
clumps, shells or filaments. The warmer of those ($T\simeq 10^4$K) produce optical emission lines,
and cooler (and denser) clumps may exist in neutral or even molecular form. Thus, the physical
conditions in quasar-driven winds are qualitatively similar to those in starburst-driven winds,
where the different phases are directly observable and spatially resolved in some nearby galaxies
\citep{heck90, veil94}. It is at present not known which of the phases of the quasar winds carries
most of the mass, momentum and energy; obtaining these measurements is critical for understanding
the full impact of quasars on galaxy formation.

The hot volume-filling component is particularly elusive because of its extremely low density, which
implies that any emission associated with it would be very weak \citep{gree14a}. The
Sunyaev-Zeldovich (SZ) effect \citep{suny70} offers a unique opportunity to detect and characterize
the bulk of quasar-driven winds. The SZ effect manifests itself as a spectral distortion of the
cosmic microwave background (CMB) that occurs when CMB photons inverse-Compton scatter off
intervening hot ionized gas. Most importantly, the magnitude of the total integrated SZ effect is
proportional to the total thermal energy of the hot ionized gas and thus allows us to measure the
thermalized energetic output from the AGN, which is not detectable by other methods. Furthermore,
the surface brightness of the SZ effect is not subject to dimming with redshift, facilitating the
study of high-$z$ systems of low-density with ionized gas that is too faint to be studied in
emission with current facilities.

Fortunately, measurements of the SZ effect have recently reached a maturity that enables
investigations of AGN feedback. Although the SZ effect was theorized over 40 years ago as a CMB
spectral distortion by hot gas around galaxy clusters, only within the past 5 years have
measurements of the SZ effect begun to reach their potential with detections of hundreds of clusters
in millimetre-wave surveys by the Atacama Cosmology Telescope (ACT), the South Pole Telescope (SPT)
and the \emph{Planck} satellite \citep{hass13b,blee15,plan15SZ}. The SZ effect associated with the
ionized circumgalactic medium of individual galaxies or galaxy groups is generally too weak to be
directly detected in these surveys. However, the large area of these surveys has enabled
measurements of the average SZ effect associated with ensembles of these lower mass systems selected
from overlapping optical surveys \citep{hand11, plan13LBGSZ, grec15, spac16}. These measurements and
associated hydrodynamic simulations \citep{batt12, lebr15} consider the effects of AGN feedback
processes on the mean gas distribution and the scaling of the SZ effect with halo mass. The
aforementioned studies, however, do not distinguish systems with ongoing AGN feedback from those
without, focusing instead on the bulk sample with mass as the defining independent variable.

The SZ effect in systems with ongoing feedback has been forecast through both analytic estimates
\citep{yama99,nata99,plat02,pfro05,chat07} and simulations
\citep{chat08,scan08,prok10,prok12,cen15b}. As discussed above, however, key details of the energy
transfer mechanism (and therefore the source of the SZ effect) are not known and will depend on the
characteristics of the system (e.g., radio-loud versus radio-quiet). Therefore measurements are
essential for progress. Given the angular resolution and sensitivity of present millimetre-wave
surveys, these measurements are limited to integrated measurements of the SZ effect over the entire
volume of the circumgalactic medium. With this single integrated measurement, the SZ effect due to
AGN feedback will only be distinguishable from an active system's \textit{in situ} SZ effect from
gravitational heating (virialization) if the feedback significantly alters the internal thermal
energy of the system \emph{during the active phase}. For radio-loud AGN, \citet{gral14} found that
the internal thermal energy due to virial equilibrium could explain the observed SZ effect, implying
that energy from AGN feedback did not significantly alter the total thermal energy. The average halo
mass of radio-loud AGN ($\gtrsim 10^{13}h^{-1}$~M$_\odot$) is found to be an order of magnitude
greater than that of radio-quiet quasars in optical lensing studies at redshifts $z<0.3$
\citep{mand09}. Thus the SZ effect from virialized gas associated with radio-quiet quasars at these
redshifts should be one to two orders of magnitude less than that of their radio-loud counterparts.
Therefore, one may hope to distinguish the SZ effect due to feedback in these lower mass systems
from that due to virial equilibrium. However, the characteristic mass of quasars at high redshift is
less constrained, with estimates in the range $(1-5)\times 10^{12}h^{-1}$M$_\odot$
\citep{whit12,shen13,wang15} and up to $10^{13}h^{-1}$~M$_\odot$ \citep{rich12}. Progress towards
understanding feedback in these high-$z$ quasars is therefore limited by these uncertainties on
their mass.

Initial studies of the SZ-effect in quasars from the Sloan Digital Sky Survey (SDSS) have used data
from the \emph{Wilkinson Microwave Anisotropy Probe} \citep{chat10} and the \emph{Planck} satellite
\citep{ruan15}. In these studies, in \citet{gral14}, and in this work, a major challenge is
estimating the SZ effect in the presence of emission from the AGN and its host. In radio-selected
systems, as used in \citet{gral14}, and at lower frequencies, as used in \citet{chat10}, the
synchrotron emission from electrons accelerated by the AGN must be modelled. At higher frequencies
dust becomes an important factor for all quasar hosts, but especially those with substantial ongoing
star formation rates. (Radio-selected systems live primarily in massive elliptical systems with
lower rates.) Different approaches, including spectral energy distribution (SED) modelling
\citep{chat10,gral14} and internal linear combination methods \citep{ruan15}, have been used to
disentangle the SZ effect from the emission component. This study uses SED modelling as in
\citet{gral14} but with the important advantage of having redshift measurements for all the systems.
We will return to consider the previous studies and energetics as measured by the SZ effect in the
discussion of our results.

In this paper, we statistically search for the hot haloes around quasars with the SZ effect in a
large sample of radio-quiet quasars from the SDSS \citep{york00, schn10, eise11, pari14} using
millimetre/submillimetre data from the ACT and \emph{Herschel}-SPIRE\@. We describe the data in
Section~\ref{sec:data} and our stacking analysis in Section~\ref{sec:stacking}. We discuss our
modelling technique and its results in Section~\ref{sec:modelling} and the implications of these
results in Section~\ref{sec:discussion}. We conclude in Section~\ref{sec:conclusions}. Unless
specified otherwise, we assume a flat $\Lambda$ cold dark matter cosmology with
$\Omega_{\rm M}=0.30$ and $\Omega_{\Lambda}=0.70$. The Hubble constant is parametrized as
$H_0=70 h_{70}$~km s$^{-1}$ Mpc$^{-1}$ and if no factor is stated $h_{70} = 1$ has been used. For
halo mass values quoted from the literature, we use the conventional unit of $h^{-1}{\rm M}_\odot$
where $h = H_0/(100$~km s$^{-1}$ Mpc$^{-1}$). The function
$E(z)=\sqrt{\Omega_{\rm M}(1+z)^3+\Omega_{\Lambda}}$ describes the evolution of the Hubble
parameter, $H(z)=H_0E(z)$.

\section{Data}
\label{sec:data}
\subsection{ACT Data}
ACT is a six-metre survey telescope for millimetre-wave astronomy operating in the Atacama Desert
high in the Andes of northern Chile \citep{fowl07, swet11}. Since 2007, ACT has surveyed more than
1000 square degrees of sky at declinations both south of and on the celestial equator. The
reduction, calibration and beam determination for the ACT data are described in \citet{dunn13},
\citet{haji11}, and \citet{hass13a}. During the years 2007--2010, ACT was equipped with the
Millimetre Bolometric Array Camera (MBAC), which operated at 148, 218, and 277~GHz, frequencies well
suited for studying the CMB, the SZ effect, and thermal dust emission. In 2013, ACT was equipped
with a new polarization-sensitive receiver and became ACTPol with a revised frequency coverage
focused on 90 and 150~GHz \citep{niem10}. The data used in this study derive from the original ACT
observations obtained on the celestial equator from 2008 to 2010. This three frequency (148, 218,
and 277~GHz) data set is the same as used in the study of millimetre emission and the SZ effect in
radio-loud AGN of \citet{gral14}. The survey area is approximately 340 square degrees
($-55\degr \leq \alpha \leq 58\degr$, $-1\fdg5 \leq \delta \leq 1\fdg5$). In
Table~\ref{tab:data_properties} we summarize the properties of the ACT data. The instrument beams
were measured using Saturn \citep{hass13a}. The 148~GHz data were calibrated through
cross-correlation with \textit{WMAP} \citep{haji11}. The 148~GHz calibration was transferred through
cross-correlation to the 218~GHz data. As a result, the uncertainties in flux calibration of the
148 and 218~GHz data are correlated at the level of 60\%. The 277~GHz data are calibrated from
Uranus \citep{hass13a}. Flux densities were estimated from a matched filtered map as described in
\citet{mars14}. The flux calibration uncertainties in Table~\ref{tab:data_properties} include
conservative estimates of error due to uncertainty in beam solid angle, absolute calibration and
map-making flux density recovery. We refer the reader to \cite{gral14} for more details. ACT is well
suited to this study due to its frequency coverage, which spans the SZ spectrum, and its small beam
size. ACT's arcminute scale resolution provides an advantage in sensitivity to the stacked quasar
signal (over e.g. \textit{WMAP} and \textit{Planck}) while maintaining a large survey area allowing
for sufficient overlap with existing surveys and catalogues.

\subsection{\herschel Data}
The {\it Herschel Space Observatory}~\/Spectral and Photometric Imaging REceiver (SPIRE) observed at
frequencies of 600, 857 and 1200~GHz \citep{grif10}. For this study we use the publicly available
\textit{Herschel} Stripe 82 Survey
\citep[HerS;][]{vier14}.\footnote{www.astro.caltech.edu/hers/HerS\_Home.html} HerS covers 79 square
degrees overlapping the celestial equator ($13\degr < \alpha < 37\degr$,
$-2\degr < \alpha < 2\degr$). The properties of HerS are listed in Table~\ref{tab:data_properties}.
The SPIRE flux calibration uncertainty and beam full width at half-maximum (FWHM) are derived from
observations of Neptune \citep{grif13}. A significant fraction of the typical rms noise values for
\herschel flux recovery, as quoted in Table~\ref{tab:data_properties}, originates from confusion
noise of the order of 8~mJy beam$^{-1}$ \citep{vier14}.

\begin{table*}
\caption{ACT and HerS survey parameters.}
\begin{center}
\begin{threeparttable}
\begin{tabular}{ccccccc}
\hline
 & \multicolumn{3}{c}{ACT} & \multicolumn{3}{c}{HerS} \\
 Band & 148~GHz & 218~GHz & 277~GHz & 600~GHz & 857~GHz & 1200~GHz \\  \hline
 Beam FWHM & 1.4~arcmin & 1.0~arcmin & 0.9~arcmin & 0.6~arcmin & 0.4~arcmin & 0.3~arcmin \\
 Typical rms noise (mJy beam$^{-1}$) & 2.2 & 3.3 & 6.5 & 13.3 &  12.9 & 14.8 \\
 Flux calibration uncertainty (per cent)  & 3\tnote{$a$} & 5\tnote{$a$} & 15  & 7\tnote{$b$} & 7\tnote{$b$} & 7\tnote{$b$} \\ \hline
\end{tabular}
\begin{tablenotes}
\item[$a$] A correlated component to the calibration uncertainty of 3 per cent between the ACT 148
  and 218~GHz bands is also accounted for.
\item[$b$] The calibration uncertainty of the \herschel data includes a 5 per cent component
  correlated across all three bands.
\end{tablenotes}
\end{threeparttable}
\end{center}
\label{tab:data_properties}
\end{table*}

\subsection{SDSS Quasar Catalogue}
\label{sec:qso_cat}
For this study we use the SDSS optical quasar catalogues derived from the spectroscopic quasar
samples in Data Release 7 \citep[DR7;][]{schn10} and Data Release 10 \citep[DR10;][]{pari14}. We
select all quasars within the ACT equatorial survey. Any object present in both DR7 and DR10 is
included only once. This combined catalogue spans a wide range of redshift out to $z \sim 7$. As we
will be performing a statistical analysis on this sample, we cut the catalogue to lie within the
well populated redshift range $0.5 < z < 3.5$, where 95\% of the quasars are found. Due to the
statistical nature of this work, the primary benefit of using a spectroscopic catalogue is not the
redshift precision it provides but the robust identification of objects as quasars which the SDSS
spectroscopic classification pipeline enables. We additionally cut this catalogue by excising all
quasars that lie within $2.5$~arcmin of sources detected with significance $> 5\sigma$ in the ACT
millimetre data. Detected ACT sources are typically extremely luminous blazars, local ($z \ll 1$)
star-forming galaxies, or high-$z$ lensed dusty star-forming galaxies \citep{mars14}. Due to their
high millimetre fluxes and the manner in which the ACT maps are filtered, these objects may
significantly contaminate the measured fluxes of nearby quasars. However, we find that neglecting
this cut alters the stacked fluxes (see Section~\ref{sec:stacking}) by no more than $1\sigma$, and
extending the mask to 5~arcmin around detected sources produces a change in the stacked signal
much smaller than the statistical uncertainties on these data.

\begin{figure}
\centering
\includegraphics[width=\columnwidth]{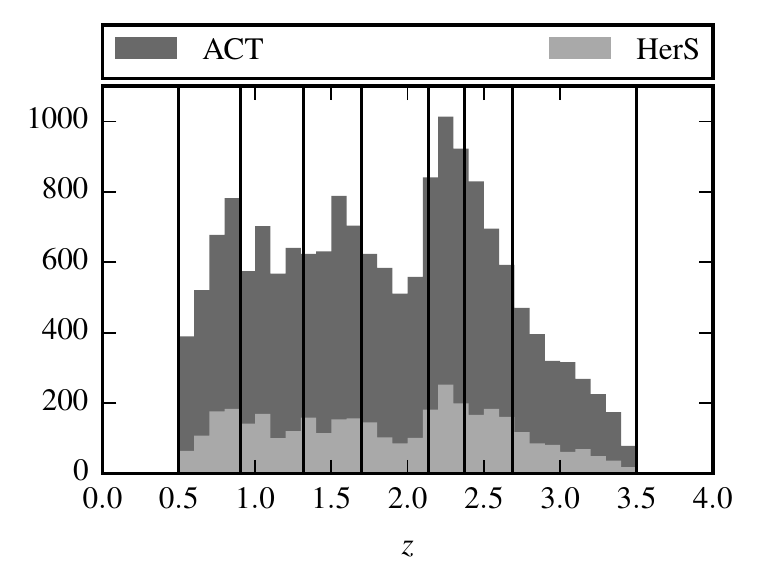}
\caption{The redshift distribution of the combined SDSS DR7 and DR10 spectroscopic quasar sample
  falling within the ACT equatorial region ($-55\degr \leq \alpha \leq 58\degr$,
  $-1\fdg5 \leq \delta \leq 1\fdg5$) after the radio loud cut has been applied. The histogram
  of the entirety of this sample is shown in dark grey and the subset which lies within the HerS
  region is shown in light grey. The vertical black lines denote the boundaries of the redshift bins
  we use for the stacking analysis. \label{fig:z_dist}}
\end{figure}

ACT galaxy clusters detected through the SZ effect are typically found at moderate significance and,
with redshifts $z \lesssim1$, should not be strongly correlated with the quasar catalogue or
contaminate our results. We therefore do not excise these objects. The catalogue obtained after
these cuts contains 17468 quasars (8642 from DR7 and 8826 from DR10). A subset of 3833 of these
objects additionally overlap with the HerS region. Fig.~\ref{fig:z_dist} shows the redshift
distribution of this quasar sample as well as the redshift distribution of those quasars that
additionally overlap with the \herschel HerS region. The sharp increase in sources at $z > 2$ shown
in Fig.~\ref{fig:z_dist} is due to the BOSS selection in DR10. The non-uniformity of this sample's
selection is mitigated in our results by binning in redshift as described in
Section~\ref{sec:stacking}. We additionally construct estimates for the optical bolometric
luminosity of each quasar in the sample by applying the bolometric correction from \citet{rich06a}
as described in Appendix~\ref{apsec:lum_det}.

\subsection{Radio Loud Cut}
\label{subsec:rl_lbol}
Since we are interested in detecting the SZ effect, which is most prominent at millimetre
wavelengths, we further cut this catalogue to exclude potentially radio-loud quasars. This acts to
minimize contamination of the SZ signal from synchrotron emission. We perform a radio-quiet
classification based on the methods of \citet{xu99}. This study finds a bimodal distribution in the
ratio of AGN radio luminosity at 5~GHz and [OIII] 5007~{\AA} line luminosity (an orientation
insensitive measure of the AGN intrinsic luminosity) and defines a cut based on this bimodality. We
use the mean relation given in \citet{reye08} between $\mathcal{M}_{2500}$, the absolute magnitude
at the rest-frame wavelength of 2500~{\AA}, and $L_{\rm [OIII]}$. Here, as in \citet{reye08}, we use
the SDSS $i$ band absolute magnitude $\mathcal{M}_I (z=2)$ \citep{rich06b} as a proxy for
$\mathcal{M}_{2500}$, which is valid for $z \sim 2$. To determine the radio luminosities, we use the
matched 1.4~GHz fluxes obtained from the Faint Images of the Radio Sky at Twenty-cm
\citep[FIRST;][]{beck95} survey that are included in the DR7 and DR10 quasar catalogues. These 1.4~GHz
radio fluxes are related to $L_{\rm 5GHz}$ using the catalogue redshifts and assuming a synchrotron
spectrum, $L_\nu \propto \nu^{-1}$. This radio-loud cut shifts the stacked fluxes (as described in
Section~\ref{sec:stacking}) in the 148 and 218~GHz bands by at most 1$\sigma$ in any redshift bin
and reduces the sample size by 412 (2\%). This cut removes the majority of sources that are detected
by the FIRST survey. Applying stricter cuts on the radio luminosities has a negligible effect on our
results.

\section{Stacking Analysis}
\label{sec:stacking}
The typical submillimetre flux densities of optically selected quasars are found to be significantly
less than the detection limits of the ACT and \herschel data. We therefore employ a stacking
analysis to measure the ensemble flux densities associated with the quasars. Stacking can be shown
to be an unbiased maximum likelihood estimator of the mean flux density of a catalogue in both the
case of a Gaussian noise background, as is the case for the point source-filtered ACT data, or for
the confusion-limited \herschel data \citep{mars09, vier13}.

Due to both the non-uniform selection of our sample in redshift and the strong redshift dependence
of our estimated bolometric luminosities (See Figs~\ref{fig:z_dist} and~\ref{fig:lbol_z}) we choose
to bin the quasar catalogue by redshift prior to stacking. This additionally allows us to account for
the redshift dependence of the parameters of the models to be fit. We choose these bins such that
they each contain approximately equal numbers of sources. For the seven redshift bins thus chosen,
we have approximately 2430 quasars per bin. The boundaries of these bins are shown as the vertical
lines in Figs.~\ref{fig:z_dist} and~\ref{fig:lbol_z}. As the catalogue was selected to lie within the
ACT equatorial region all of these objects contribute to the stacks for the 148, 218 and 277~GHz
bands. For the 600, 857 and 1200~GHz \herschel bands the number of quasars that contribute is
further diminished by the band dependent masks which account for varying sky coverage of the data.
The smaller sky coverage of the HeRS survey results in 506--574 quasars per bin with submillimetre
data.

\begin{figure*}
  \includegraphics[width=\textwidth]{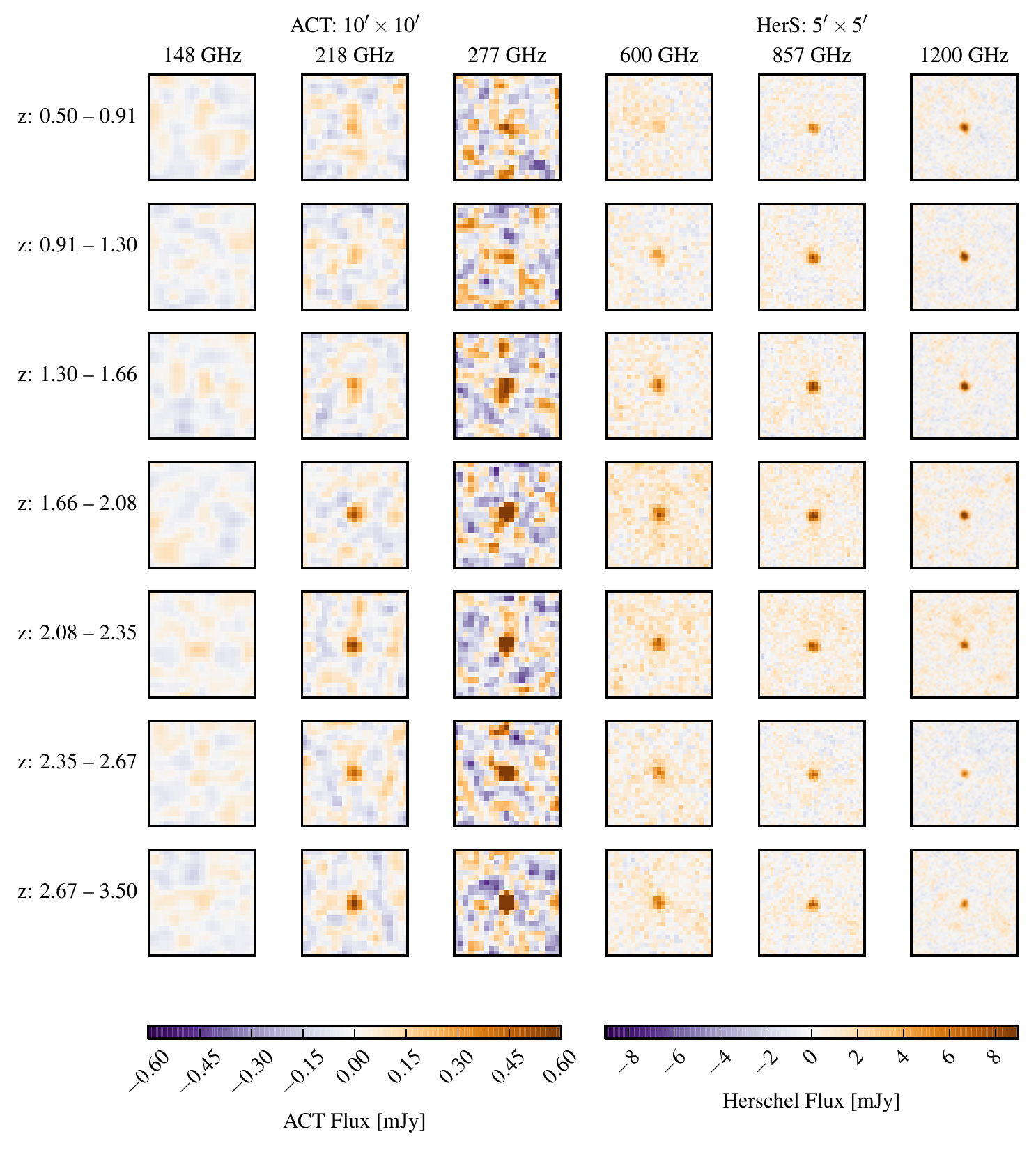}
  \caption{
   Weighted average thumbnail images around the quasar positions in 10~arcmin $\times$ 10~arcmin stamps for the matched filtered ACT data (left three columns) and 5~arcmin $\times$ 5~arcmin stamps for the \herschel data (right three columns). The observed signal is dominated
   by the dust emission associated with quasars. Since no background
   aperture has been subtracted from the \herschel data here, large scale excess positive
   signal, associated with correlated cosmic infrared background emission, is visible.
   \label{fig:stacks}}
\end{figure*}

For each of the redshift bins, we construct the stacked signal in a band by taking the inverse
variance weighted average of the measured flux density corresponding to each source in the bin:
\begin{equation}
  \label{eq:stack}
  \bar{d}_\nu^\alpha = \frac{\sum_i^{N^\alpha} w_{i,\nu} d_{i,\nu}}{\sum_i^{N^\alpha} w_{i,\nu}}.
\end{equation}
This yields the stacked signal $\bar{d}_\nu^\alpha$ of the redshift bin $\alpha$ containing
$N^\alpha$ sources, each contributing flux density $d_{i,\nu}$ in the band corresponding to a
frequency $\nu$. The inverse variance weights $w_{i,\nu}$, which vary from source to source and
between bands, are determined from the error maps for each data set. For the \herschel data, these
error maps are derived from the diagonal component of the pixel-pixel covariance matrices that are a
product of the map making pipeline \citep{pata08}. For the matched filtered ACT data, we take the
variance in a pixel for a given band to be inversely proportional to the value of that band's
hits-count map at that location. The flux density contributions $d_{i,\nu}$ of each source are taken
from the pixel values at the  location of each object in our catalogue. These values are corrected for
the averaging effects of each map's pixelization scheme.

The uncertainties on the stacked flux densities, $\sigma_\nu^\alpha$ ,are determined through a
bootstrap analysis which resamples, with replacement, the sources contributing to each bin. We then
calculate the uncertainties by taking the standard deviation of the stacked flux densities measured
for each of the resamples. The bootstrapped uncertainties thus determined are found to be within
10\% of the standard errors on the mean of each of the binned catalogues. Table~\ref{tab:stack_table}
shows the resulting stacked flux densities in each band and redshift bin. Thumbnail images of each
of these stacks are shown in Fig.~\ref{fig:stacks}.

\begin{table*}
  \caption{Stacked quasar flux densities for the ACT (148--277~GHz) and \herschel (600--1200~GHz) data.}
\centering
\begin{threeparttable}
\begin{tabular}{ccccccccc}
  \hline
  Bin &
$z$ range &
$N^{\alpha}$ \tnote{$\dagger$} &
$\bar{d}_{\rm 148~GHz}^\alpha$ &
$\bar{d}_{\rm 218~GHz}^\alpha$ &
$\bar{d}_{\rm 277~GHz}^\alpha$ &
$\bar{d}_{\rm 600~GHz}^\alpha$ &
$\bar{d}_{\rm 857~GHz}^\alpha$ &
$\bar{d}_{\rm  1200~GHz}^\alpha$ \\
& & & (mJy) & (mJy) & (mJy) & (mJy) & (mJy) & (mJy) \\
  \hline
   1 & 0.50--0.91 & 2432 (545) & 0.06 $\pm$ 0.04 & 0.26 $\pm$ 0.07 & 0.6 $\pm$ 0.16 & 1.4 $\pm$ 0.8 & 5.7 $\pm$ 0.7 & 10.2 $\pm$ 0.8 \\ 
 2 & 0.91--1.30 & 2435 (518) & 0.03 $\pm$ 0.04 & 0.22 $\pm$ 0.07 & 0.4 $\pm$ 0.16 & 4.0 $\pm$ 0.9 & 7.2 $\pm$ 0.8 & 10.0 $\pm$ 0.9 \\ 
 3 & 1.30--1.66 & 2431 (506) & 0.09 $\pm$ 0.04 & 0.33 $\pm$ 0.07 & 0.8 $\pm$ 0.16 & 5.9 $\pm$ 0.9 & 9.9 $\pm$ 0.9 & 13.1 $\pm$ 1.0 \\ 
 4 & 1.66--2.08 & 2434 (482) & 0.05 $\pm$ 0.04 & 0.51 $\pm$ 0.07 & 1.3 $\pm$ 0.16 & 7.6 $\pm$ 0.9 & 9.5 $\pm$ 0.8 & 10.5 $\pm$ 0.9 \\ 
 5 & 2.08--2.35 & 2432 (546) & 0.11 $\pm$ 0.04 & 0.57 $\pm$ 0.07 & 1.3 $\pm$ 0.17 & 5.0 $\pm$ 0.8 & 6.9 $\pm$ 0.8 & 6.8 $\pm$ 0.7 \\ 
 6 & 2.35--2.67 & 2436 (574) & 0.07 $\pm$ 0.04 & 0.44 $\pm$ 0.07 & 1.2 $\pm$ 0.17 & 4.5 $\pm$ 0.7 & 6.2 $\pm$ 0.7 & 6.6 $\pm$ 0.7 \\ 
 7 & 2.67--3.50 & 2434 (564) & 0.12 $\pm$ 0.05 & 0.59 $\pm$ 0.07 & 1.8 $\pm$ 0.17 & 5.1 $\pm$ 0.7 & 7.3 $\pm$ 0.7 & 6.5 $\pm$ 0.8 \\ 

  \hline
\end{tabular}
\begin{tablenotes}
  \item[$\dagger$] Numbers in parentheses are the bin counts which overlap the HerS region.
\end{tablenotes}
\end{threeparttable}
\label{tab:stack_table}
\end{table*}

We perform null tests on the stacked signal by randomly drawing catalogues using positions uniformly
distributed over the sky coverage of the ACT equatorial data. The stacking procedure is otherwise
left unchanged with identical masks applied, producing similar numbers of null sources with
\herschel data as we have in our quasar catalogue. We run 500 of these null stacks, each using a
catalogue of the same size as a single one of our redshift bins. The mean stacked fluxes of these
null tests are found to be consistent with zero with no detectable bias across all bands. Testing
against the hypothesis of no stacked signal, we obtain a $\chi^2$ = 6.2 with 6 degrees of freedom
(one for each band) yielding a probability to exceed (PTE) of 0.40.

Stacking analyses are subject to bias undetectable through null tests if there exist sources in the
data that are significantly correlated with the catalogue that is being used for the stack.
\citet{wang15} have detected a correlation between SDSS optically selected quasars and the cosmic
infrared background over the same redshift range as the sample we are studying. To correct for this
in the \herschel data we perform an aperture photometry background subtraction using the mean flux
in a tight annular aperture with inner and outer diameters of 2 and 3.2 beam FWHM, respectively, for
each of the 600, 857 and 1200~GHz bands. This subtraction amounts to an average correction of
$\lesssim 1$~mJy (10--25\%) for the \herschel stacked fluxes. An analogous background subtraction is
effectively provided by the matched filter applied to the ACT data.

\section{Modelling}
\label{sec:modelling}
\subsection{Constructing Stacked Models of the Quasar SED}
\label{subsec:stacked_model}
Our model for the stacked quasar spectral energy distribution (SED) has two components: a greybody
dust spectrum and a contribution from the SZ effect. We parameterize the mean quasar dust spectrum in terms of the
rest frame frequency $\nu$ as
\begin{equation}
  \label{eq:1}
  S_{\rm dust}(\nu, z, L_{\rm ir}, \beta, T_{\rm d}) = \frac{L_{\rm ir}}{4\pi D_{\rm L}^2(z)} \frac{((1 +
    z)\nu)^\beta B_{(1+z)\nu}(T_{\rm d})}{\int \nu^{\prime\beta} B_{\nu^\prime}(T_{\rm d}) \mathrm{d}\nu^\prime},
\end{equation}
representing a modified (optically thin) blackbody spectrum at a temperature $T_{\rm d}$ with
emissivity $\beta$. The integral in the denominator is taken from 300 to 21~THz (14--1000~\micron)
in the rest frame such that $L_{\rm ir}$ is representative of the infrared bolometric luminosity of
the quasar. We use the quasar's redshift and corresponding luminosity distance, $D_{\rm L}(z)$, to
derive the model SED in terms of the observed frame flux density. We discuss variations on this dust
model in Appendix~\ref{apsec:alt_models}.

The contribution of the SZ effect is parameterized in terms of the volume integrated thermal
pressure of the electron gas, $\pdv = {\smallint}n_{\rm e} T_{\rm e}\mathrm{d}V$. Integrated over
the solid angle of a source, the SZ effect makes a contribution to the observed flux density with
the form
\begin{equation}
  \label{eq:sz_definition}
  S_{\rm SZ}(\nu, z, {\smallint}p\mathrm{d}V) = I_0 g(\nu) \frac{\sigma_{\rm T}}{m_{\rm e} c^2}
  \frac{{\smallint}p\mathrm{d}V}{D^2_{\rm A}(z)}.
\end{equation}
Here, $I_0 = 2 (k_{\rm B} T_{\rm CMB})^3/(hc)^2$ and the SZ spectral function is
\begin{equation}
  \label{eq:6}
  g(x \equiv h\nu/k_{\rm B} T_{\rm CMB}) = \frac{x^4{\rm e}^x}{({\rm e}^x - 1)^2} \left(x \frac{{\rm e}^x+1}{{\rm e}^x-1} - 4\right),
\end{equation}
the non-relativistic form of the SZ spectrum. With the levels of SZ measured in this work, the
associated temperature for a thermalized medium is approximately 1~keV implying relativistic
corrections of 1--2 per cent \citep{fabb81, reph95}. Additional SZ signal could originate from a
smaller, hot relativistic plasma that is not in thermal equilibrium with the larger circumgalactic
medium, however, the sensitivity and spectral resolution of our measurements do not allow us to
explore this possibility.

The quantity $\pdv$ used here represents only the volume integrated thermal pressure of the electron
gas, which can be related to the total thermal energy through
\begin{equation}
  \label{eq:etherm}
  E_{\rm th} = \frac{3}{2}\left(1 + \frac{1}{\mu_{\rm e}}\right)\pdv,
\end{equation}
where we take the mean molecular weight per free electron to be $\mu_{\rm e} = 1.14$.

These dust and SZ contributions are combined to form the full model SED for each quasar,
\begin{equation}
  \label{eqn:sed_tot}
  S_{\rm tot} = S_{\rm dust}(\nu, z, \beta, L_{\rm ir}, T_{\rm d}) + S_{\rm SZ}(\nu, z, {\smallint}p\mathrm{d}V).
\end{equation}

For each quasar $i$ in a redshift bin $\alpha$ we use the inverse-variance weights
from the data to construct the stacked model SED,
\begin{equation}
  \label{eq:3}
  \bar{S}_\nu^\alpha = \frac{\sum_i^{N^\alpha} w_{i,\nu}~S_{\rm tot}(\nu, z_i, {\smallint}p\mathrm{d}V_i, \beta, L_{\rm ir}^\alpha, T_{\rm d})}{\sum_i^{N^\alpha} w_{i,\nu}}.
\end{equation}

The greybody parameters may exhibit significant redshift evolution. We fit for $L_{\rm ir}$
independently in each redshift bin while $T_{\rm d}$ and $\beta$ are fit globally, each with a
single value across all the redshift bins. We find that allowing $T_{\rm d}$ to vary independently
across the bins results in consistent temperatures for all bins and tends to produce $\chi^2$ values
which are indicative of overfitting. We address other approaches for modelling dust in
Appendix~\ref{apsec:alt_models}.

We use two approaches for modelling the SZ effect. The first approach uses ${\smallint}p\mathrm{d}V$ directly
as a parameter, with a single value fit across all of the redshift bins such that ${\smallint}p\mathrm{d}V_i$
takes on the same value for each quasar. This method produces an estimate for the average thermal
energy in ionized gas associated with quasars across our entire redshift range. This model will be
referred to as the ``${\smallint}p\mathrm{d}V$ model.'' Our second approach is motivated by the hypothesis
that such a signal would be dominated by energy injected into the surrounding medium by quasar
feedback. In this scenario, assuming that cooling is negligible over the time-scales in question,
the thermal energy output from quasar feedback is
\begin{equation}
  \label{eq:4}
  E_{{\rm th}, i} = f L_{{\rm bol}, i} \tau.
\end{equation}
Here $L_{{\rm bol}, i}$ is the optically-derived bolometric luminosity of the $i$th quasar
(determined as described in Appendix~\ref{apsec:lum_det}) and $f$ is the efficiency with which this
radiative energy is able to thermally couple to the surrounding gas over the period of active quasar
activity prior to observation, $\tau$. We then fit the efficiency $f$ as a parameter in the model
and use this to scale the value of ${\smallint}p\mathrm{d}V_i$ used to generate the SED for each quasar
(given its $L_{{\rm bol}, i}$) in Equation~\ref{eqn:sed_tot} by making use of
Equation~\ref{eq:etherm}. As this efficiency is degenerate with $\tau$ in determining
${\smallint}p\mathrm{d}V_i$, we normalize $\tau$ to a fiducial active period of
$\tau = \tau_8 \times 10^{8}$~yr and report our value for the efficiency $f$ in units of
$\tau_8^{-1}$~per cent. We refer to this model as the ``quasar feedback'' model.

\subsection{Results of Modelling the Quasar SED}
\label{subsec:modelling_results}
To constrain the parameters in these models, we construct a Gaussian likelihood function,
\begin{equation}
  \label{eq:5}
  \ln{\mathcal{L}}  = -\frac{1}{2} \sum_\alpha^{N_{\rm bins}} (\bm{d}^\alpha - \bm{S}^\alpha)^{\rm T} \textsf{\textbf{C}}_\alpha^{-1} (\bm{d}^\alpha - \bm{S}^\alpha).
\end{equation}

Here, $\bm{d}^\alpha$ an $\bm{S}^\alpha$ are vectors of the stacked data and model fluxes
respectively with each element corresponding to a separate band. The covariance matrices
$\textsf{\textbf{C}}_\alpha$ are determined by combining the bootstrapped stacked flux
uncertainties, $\sigma_\nu^\alpha$, with the calibration uncertainties and covariances for the
different bands discussed in Section~\ref{sec:data}.

This likelihood is then maximized using an affine invariant Markov chain Monte Carlo (MCMC) ensemble
sampler algorithm \citep{fore13}. We choose to use unbounded uniform priors for the parameters,
$L_{\rm ir}^\alpha$ and $T_{\rm d}$. For the parameters $\beta$, ${\smallint}p\mathrm{d}V$ and $f$
we again use uniform priors but restrict these parameter values to be $> 0$ to avoid unphysical
degeneracies. We test for convergence of the chains to the posterior distribution by evaluating
their autocorrelation times relative to the length of the sampled chain. All best-fitting parameter
constraints are reported as their 50th percentiles when marginalized over all other parameters in
the posterior distribution. The uncertainties are determined by the marginalized 68 per cent
credible regions around these values. We calculate $\chi^2$ values as
$-2 \ln{\mathcal{L}}_{\rm max}$ where $\ln{\mathcal{L}}_{\rm max}$ is calculated with a
conjugate-gradient optimization algorithm initialized at the best-fitting parameter locations from
the MCMC chains.

\begin{figure*}
  \includegraphics{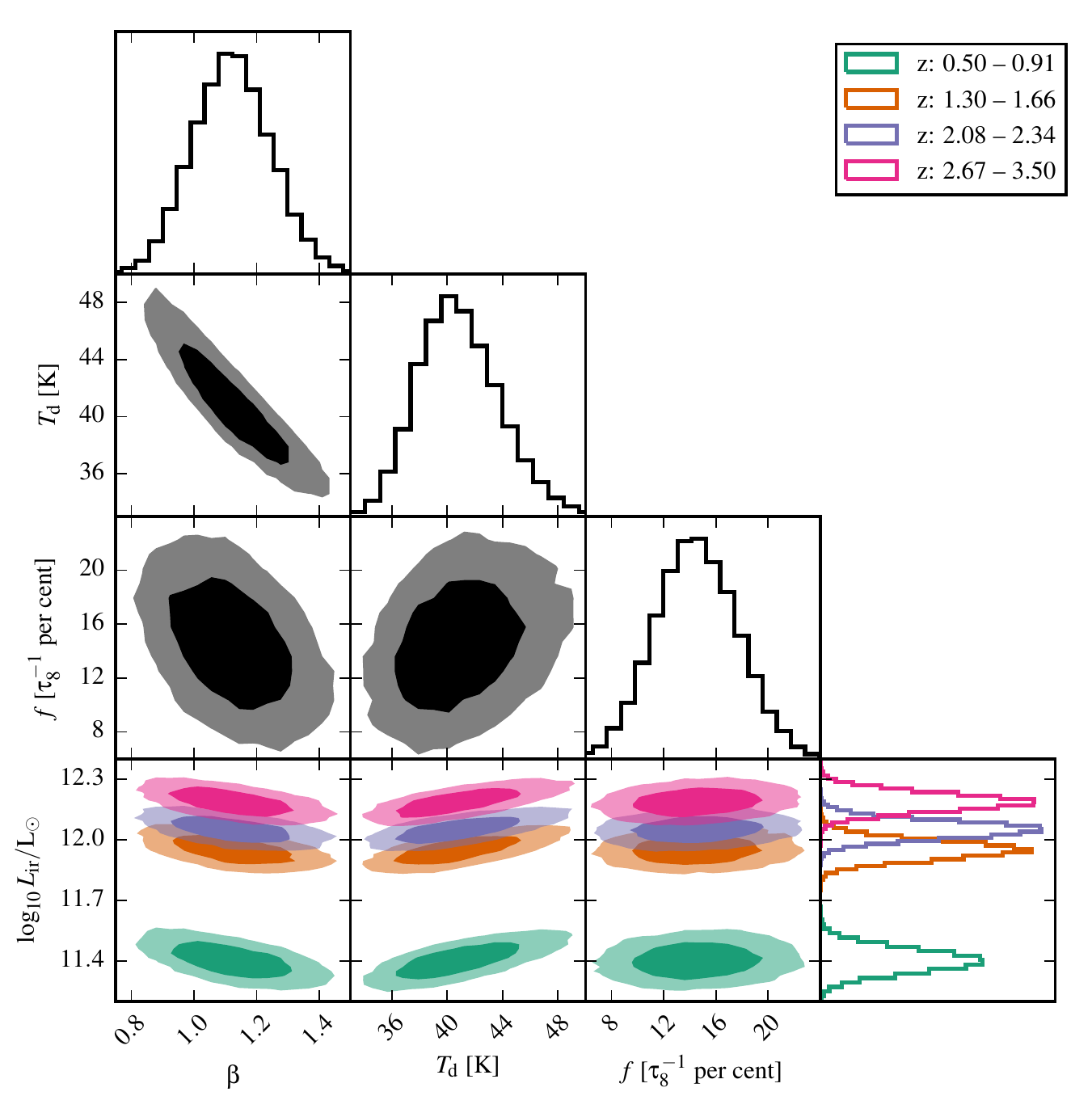}
  \caption{Contours of the 68 and 95 per cent credible intervals of the parameters for the
    quasar efficiency model in light and dark shades respectively. The histograms display each
    parameter's marginalized posterior distribution. Parameters $f$, $\beta$ and $T_{\rm d}$ are fit
    globally across all redshift bins. For the other parameters each colour indicates the
    constraints from a single redshift bin. For clarity, we show the bin dependent parameters for a
    subset of the redshift bins used.
    \label{fig:param_contours}}
\end{figure*}

\begin{table}
\caption{Marginalized parameter constraints for the quasar feedback model}
\centering
\begin{tabular}{cccccc}
  \hline
$z$ range &
$\log_{10}(L^\alpha_{\rm ir}/L_{\odot})$ &
$T_{\rm d}$ &
$\beta$ &
$f$ \\
&& (K) &&
($\tau_8^{-1}$~per cent)
\\
\hline
0.50--0.91 & $11.40 \pm 0.06$ & \multirow{6}{*}{$40.68^{+3.17}_{-2.70}$} &\multirow{6}{*}{$1.12^{+0.13}_{-0.12}$} &\multirow{6}{*}{$14.5^{+3.3}_{-3.1}$} \\
0.91--1.30 & $11.66 \pm 0.05$ &  & \\
1.30--1.66 & $11.95 \pm 0.05$ &  & \\
1.66--2.08 & $12.06 \pm 0.04$ &  & \\
2.08--2.34 & $12.05 \pm 0.04$ &  & \\
2.35--2.67 & $12.04 \pm 0.05$ &  & \\
2.67--3.50 & $12.19 \pm 0.05$ &  & \\

\hline
\end{tabular}
\label{tab:param_tab}
\end{table}

\begin{figure*}
  \includegraphics[width=\textwidth]{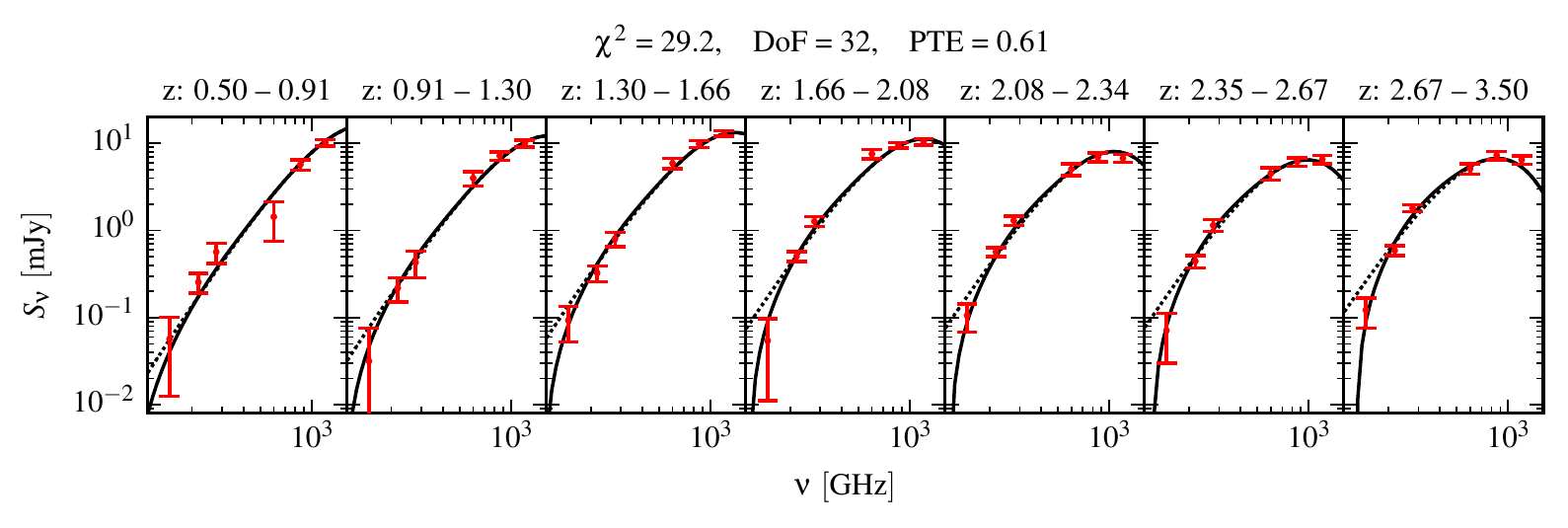}
  \includegraphics[width=\textwidth]{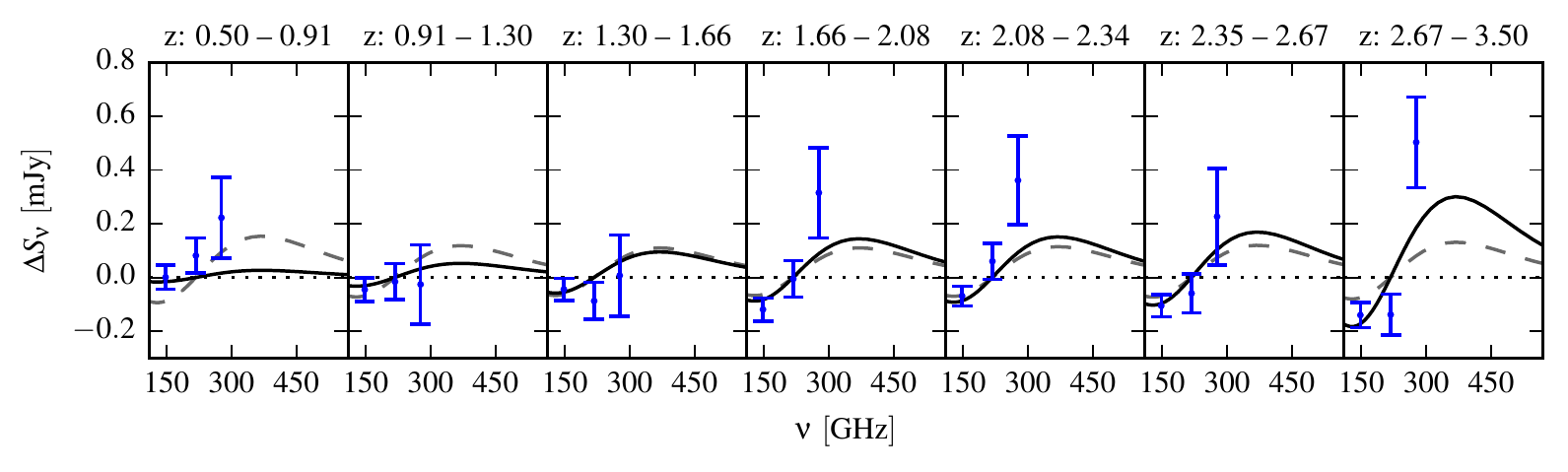}
  \caption{
    Top: stacked millimetre to infrared SEDs. Red data points represent the stacked data in
    our bands spanning 148--1200~GHz. The best-fitting stacked SEDs for the quasar feedback model for each redshift bin are shown in the
    solid black lines. The dotted black lines show the stacked model with no SZ effect.
    Bottom: dust subtracted millimetre SEDs highlighting the fit to the SZ component. The
    dust subtracted data are shown as the blue data points and the solid
    black lines represent the SZ component of the quasar feedback stacked model. The dashed grey
    lines shows the best-fitting SZ component when parameterized using the ${\smallint}p\mathrm{d}V$
    formalism. The $\Delta S_\nu = 0$ line, corresponding to a dust only signal is shown as the
    dotted black line. All errorbars shown here are statistical only and do not include the calibration
    uncertainties and covariances that are accounted for in the modelling.
    \label{fig:bin_sed}}
\end{figure*}

Using the models outlined in Section~\ref{subsec:stacked_model}, we fit the parameters
$\beta$, $L_{\rm ir}^\alpha$, $T_{\rm d}$ and either $\pdv$ or $f$ where the
$\alpha$ index labels each of our seven redshift bins.

For the $\pdv$ model, we find that the data are fit well with $\pdv = \pdvval \times 10^{60}$~erg
which corresponds to a total thermal energy of $E_{\rm th} = \pdvtotval \times 10^{60}$~erg (using
Equation~\ref{eq:etherm}). The fit yields $\chi^2$ = \pdvchisq~for 32 degrees of freedom, resulting
in a PTE of \pdvpte. For the quasar feedback model, we find $f = \fval \tau_8^{-1}$ per cent with
$\chi^2$ = \fchisq~for 32 degrees of freedom, corresponding to a PTE of \fpte. Both of these models
produce consistent best-fitting values for all the other parameters. For the quasar feedback model,
the 68 and 95 per cent credible intervals for all the fit parameters are shown in
Fig.~\ref{fig:param_contours} and their marginalized constraints are summarized in
Table~\ref{tab:param_tab}. The numbers in this table refer to the median of the marginalized
parameter posteriors and the boundaries of their 68 per cent credible intervals. The maximum
likelihood stacked SEDs from this model are shown along with the data in Fig.~\ref{fig:bin_sed}. In
the lower plot of Fig.~\ref{fig:bin_sed} we show the ACT data residuals, after subtracting the
modelled dust SED\@. Overplotted is the best-fitting modelled SZ component. Both the $\pdv$ and
quasar feedback models are shown.

These fits provide evidence for the presence of hot thermalized gas associated with
these systems manifesting itself through an SZ distortion at millimetre wavelengths. A fit to the
data explicitly neglecting the SZ component is formally worse than both of the above models with
$\chi^2$ = \nullchisq~and 33 degrees of freedom. We therefore observe a $\Delta \chi^2$ improvement
of \pdvnulldeltachisq~for the $\pdv$ model and \fnulldeltachisq~for the quasar feedback model,
both of which add one additional parameter. This corresponds to $3\sigma$ evidence for the presence
of associated thermalized gas or, assuming the quasar feedback scenario, $4\sigma$ evidence for the
thermal coupling of quasars to their surrounding medium.

We find the best-fitting models require significant evolution of $L_{\rm ir}$ across the redshift
range we probe, spanning the range $10^{11.4-12.2}~{\rm L}_{\odot}$\footnote{We correct an error in
  our previous work, \citet{gral14}, where the bolometric luminosity for the \citet{best12} sample
  reported in Table 1 in that work should be
  $\log_{10}(L_{\rm ir}/{\rm L}_\odot) = 9.7^{+0.1}_{-0.3}$.}. The implications of this measurement
are discussed in Section~\ref{sec:SFR}. Fitting for a single temperature across all redshift bins,
we find $T_{\rm d} = 40 \pm 3$~K. Allowing the dust temperatures to vary independently for each
redshift bin yields values all consistent with this globally fit temperature and shows no detectable
sign of redshift evolution.

The best fitting value of the dust emissivity index $\beta$ is found to be $1.12 \pm 0.13$. This is
on the low end of the theoretical range of $1 < \beta < 2$ as discussed further in
Section~\ref{subsec:dust_emissivity}. As this value for the emissivity index warrants further
investigation and due to the systematic dependence of our results on the chosen form of the dust
model, we explore a range of alternative dust models in Appendix~\ref{apsec:alt_models}. These
include dust models with values of $\beta$ fixed to values more commonly seen in the literature, as
well as composite two temperature models and models with an optically thick dust spectrum. We find
that a model with $\beta$ fixed to 1.6 produces an acceptable fit (PTE=0.16) to the data with
3$\sigma$ evidence for the SZ effect. A model with $\beta$ fixed to 1.8 does not fit the data well
(PTE=0.01). Motivated by increasing evidence that high redshift dusty galaxies are characterized by
an optically thick dust emission \citep{riec13, huan14}, we evaluate a model with an optically thick
dust component. This model fits the data (PTE=0.69) with $\beta=1.3\pm0.3$ and 4$\sigma$ evidence
for SZ\@. A two-temperature model adequately fits the data (PTE=0.35) with $\beta=1.4\pm0.1$ (for
both greybodies) and no SZ component. However, this model comes at the cost of significant
additional model complexity in the form of seven additional parameters.

\section{Discussion}
\label{sec:discussion}
Our data and modelling results establish that the far-infrared (FIR) and millimetre SEDs of
radio-quiet quasars are (1) dominated by dust emission and (2) require at the 3$\sigma$--4$\sigma$
level an SZ distortion for most dust emission models. Insomuch as the bolometric FIR luminosity of
quasars is the least model-dependent quantity in our analysis, we begin our discussion with
$L_{\rm ir}$, its redshift evolution and associated star formation rates in Section~\ref{sec:SFR}.
We then consider our best-fitting dust models in the context of previous studies in
Section~\ref{subsec:dust_emissivity}. For most of these models, an SZ effect is preferred. In
Section~\ref{sec:quasar_feedback}, we consider whether the level of the SZ is consistent with simple
virialization given the quasar host halo masses or whether energy injection from feedback should be
invoked. Finally in Section~\ref{sec:comparison}, we compare our findings to other studies of the SZ
effect associated with quasars and their hosts.

\subsection{Infrared Luminosity and Star Formation Rates of Quasars}
\label{sec:SFR}

Our models provide estimates of the FIR luminosity of quasars through the $L_{\rm ir}$ parameters
fit to our data in each redshift bin. The FIR emission associated with quasars originates from warm
dust that is heated by ultraviolet radiation from young stars and the quasar itself. In the absence
of detailed mid-infrared data, these contributions are difficult to separate. \citet{kirk12} attempt
to quantify the percentage contribution due to star formation in these systems, finding 21 and 56
per cent of the FIR bolometric luminosity is due to star formation at $z \sim 1$ and $z \sim 2$,
respectively. We instead estimate only upper limits on the star formation rates of quasar hosts by
assuming no infrared contribution originating from quasar emission and using the total infrared
bolometric luminosities of these systems to calculate their star formation rates. The $L_{\rm ir}$
values we report derive from the normalization of the greybody emission we model and therefore
represent only a portion of the total infrared bolometric luminosities. To estimate the total
infrared luminosity of star formation, $L^\prime_{\rm ir}$, we use the relationship between total
$L^\prime_{\rm ir}$ and rest-frame 160~\micron luminosity, $L_{160}$ from \citet{syme08}. For the
redshift range we probe, this rest-frame frequency falls within the frequency coverage of our data.
Using total $L^{\prime}_{\rm ir}$ calculated from $L_{160}$ values derived from the best-fitting SED
models for each redshift bin, we obtain average star formation rates using the relation from
\citet{bell03}. We find that the average star formation rates of quasar hosts are no more than
$\sim 60,~250$ and $ 400~{\rm M}_{\odot}/{\rm yr}$ for $z \sim $1, 2 and 3, respectively.

\subsection{Dust Emission Models}
\label{subsec:dust_emissivity}
For both parameterizations of the SZ effect, the best-fitting values of the dust emissivity $\beta$
for an optically thin greybody model are found to be, $1.12 \pm 0.13$. The turn over in the SEDs
prefer a greybody model of temperature $T_{\rm d}\approx40$~K, consistent with previous observations
\citep{beel06,dai12}. In Appendix~\ref{apsec:alt_models}, we evaluate fits with alternative dust
emission models with fixed emissivity, with optically thick emission, and with two dust components
at different temperatures. All models require a component with temperature $T_{\rm d}\approx40$~K.
However the dust emissivities of these alternative models skew higher, giving 68 per cent credible
regions with $\beta \lesssim 1.6$ and 95 per cent credible regions with $\beta \lesssim 1.8$. Given
the preference for lower dust emissivities in these results, we review below studies relevant to
this aspect of our dust models.

Millimetre-wavelength observations find $\beta$ to vary among different galaxies between the
theoretical bounds \citep[$1\le\beta\le 2$,][]{bohr83}, peaking at $\beta=1.5$ \citep{cari92,
  lise00}. \emph{Herschel} has revolutionized the study of FIR SEDs of extragalactic sources. A
common assumption for fitting \emph{Herschel} data is $\beta=1.5$ \citep{kirk12}, but because
$\beta$ is only relevant for the Rayleigh-Jeans tail of the SED (i.e., at wavelengths well beyond
that of the thermal peak) rarely does the wavelength coverage extend to long enough wavelengths to
constrain $\beta$ from unrestricted fits, which is especially problematic because of the presence of
components at a range of temperatures \citep{sun14}. Perhaps the most relevant comparison is with
other sub-millimetre observations. For example, values of $\beta=1.3$, are observed by \emph{Planck}
in local star-forming galaxies \citep{negr13}, consistent with spectral indices inferred by SPT in
their high-$z$ dust dominated source population \citep{moca13}. In a sub-millimetre study of
quasars, \citet{beel06} find $\beta=1.6\pm0.1$. These latter studies are broadly consistent with our
results.

Additionally, all the models require a dominant component with temperature $\sim40$~K. These
temperature values apply only to the low-frequency emission; in practice, the SED of quasars is
dominated by thermal components at much larger temperatures \citep{rich06b}, all the way to the dust
sublimation cutoff at $T\ga 1000$ K. All higher temperature components produce $\nu^{2+\beta}$
spectra at the wavelengths of our \herschel and ACT observations; thus, our fits only recover the
lowest temperature present in the multi-temperature distribution. While this higher temperature dust
tends to affect the Wein side of the blackbody spectrum, a colder component could change the shape
of the low-frequency spectrum. If there are such additional colder components at $T<40$ K, because
of our single-temperature fitting procedure, they will tend to reduce the apparent $\beta$
\citep{dunn01}.

Finally, there is observational evidence that the SEDs of high-$z$ dusty star forming-galaxies are
indicative of optically thick dust \citep{riec13, huan14}. The precise origin of the dust emission
of the quasar hosts is difficult to ascertain and the implicit assumption of dust optical depth
$ << 1$ in our fiducial dust models may not be justified. Allowing for an optically thick greybody
spectrum has the effect of broadening the dust peak and would manifest itself as an artificially
lower $\beta$ index when not accounted for.

\subsection{Interpretation of the Observed SZ Signal}
\label{sec:quasar_feedback}
In Section~\ref{subsec:modelling_results}, we show that the data favour (at the 3$\sigma$--4$\sigma$
level for most dust models) SED models for the stacked quasar sample with a thermal SZ contribution.
We find that scaling this signal by the quasar bolometric luminosities, as expected if this energy
is sourced by quasar feedback, provides a good fit to the data and yields a quasar feedback heating
efficiency of $f = \fval \tau_8^{-1}$ per cent. Formally, this percentage is an upper limit because
some of the thermal energy responsible for the SZ effect is due to simple virialization, which is
not included in the feedback model. However in the following we argue that most estimates of quasar
host halo masses imply that the contribution from virialization is modest. Furthermore we cannot
separately constrain the heating efficiency $f$ and the period of quasar activity $\tau$. Quasar
lifetimes are quite poorly known \citep{mart04}, but we can use theoretical and observational
estimates of quasar lifetimes from the literature to normalize our result to a fiducial value.
Earlier observational estimates favoured $\tau \la 10^8$ yr \citep{mart01, jako03, schi04, shen07,
  wors07, gonc08, trai13} while more recent ones suggest $\tau \ga 10^8$ yr \citep{marc04, dipo14,
  dipo15}, with values $\sim 10^8$ yr commonly suggested by models of galaxy formation
\citep{hopk05a, hopk05b}. As, on average, we should observe a given quasar halfway through its
active lifetime, the period of activity prior to observation should be estimated as the typical
quasar lifetime divided by two. In addition to being systematically dependent on this quasar
lifetime estimate, the efficiency values we derive also depend on the accuracy of our estimates of
the optical bolometric luminosities for the sample. The bolometric corrections used to derive these
values (see Appendix~\ref{apsec:lum_det}) are uncertain at the 40 per cent level. The luminosity
scaling of this signal is further explored in Appendix~\ref{apsec:lum_dep}.

In the literature, values of $\sim$5--7 per cent efficiency in thermalizing quasar radiative energy
into the gas immediately around the accretion system are more typically used in galaxy formation and
evolution models \citep[e.g.,][]{spri05, hopk06}. Feedback in this form and with these efficiency
values has been shown to bring both simulated and semi-analytic models of self-regulating quasar
feedback in line with the observed $M_{\rm BH}$--$\sigma$ relation and quasar luminosity function
\citep{wyit03, dima05}. However, the thermalization efficiency remains poorly known and the exact
value obtained from the simulations depends on the details of the feedback implementation
\citep{ciot01, nova11, choi12}.

While the quasar feedback model provides a compelling explanation for this SZ signal we also
consider the possibility that this signal results in whole or in part from hot gas in virial
equilibrium with the quasar host haloes, a more typical manifestation of the SZ effect such as that
found in galaxy clusters \citep[e.g.,][]{hass13b, blee15, plan15SZ}. In a study of the SZ effect
associated with radio-loud AGN, \citet{gral14} find a signal consistent with that expected from the
gravitationally shock heated ionized haloes of the massive ($\gtrsim 10^{13}$M$_\odot$) hosts of
these objects. While not precisely known, the halo masses of the quasars in our study are expected
to be significantly lower in mass. Galaxy-quasar clustering measurements at the lower limit of the
redshift range we target ($z\sim 0.5$) yield quasar halo masses of
$\sim 4\times 10^{12}h^{-1}$M$_{\odot}$ \citep{shen13} with a weak dependence of halo mass on quasar
bolometric luminosity. Quasar-quasar clustering on the other hand yields a halo mass of
$\sim 1\times 10^{12}h^{-1}$M$_{\odot}$ at $z=2.4$ with no evidence for luminosity dependence
\citep{whit12}. \citet{wang15} find that the cross-correlation signal between quasars and the CIB is
well fit by a halo model for clustering corresponding to a host halo mass of
$10^{12.36\pm0.87}h^{-1}$M$_\odot$ for DR9 quasars with a median redshift of $2.5$. The quasars in
the latter two of these clustering studies were optically selected type 1 objects in the SDSS-III
survey, and are thus comparable samples to that used in our work. Somewhat higher halo mass values
$\gtrsim 5\times 10^{12}h^{-1}$M$_\odot$ are found in clustering studies of $z \gtrsim 2$ unobscured
quasars by \citet{rich12} and CMB lensing studies such as \citet{sher12} and \citet{dipo14, dipo15}.
The latter two of these measurements are not directly comparable to our work as they are based on
infrared-selected quasars with photometric redshifts.

Taking a range of $(1-5)\times 10^{12}h^{-1}$M$_\odot$ we can estimate the expected magnitude of the
SZ signal due only to the gravitationally heated virialized reservoirs of hot gas that would be
associated with haloes of this mass. \citet{plan13LBGSZ} find that the integrated SZ signal,
\begin{equation}
  \label{eq:sz_Y}
  Y \equiv \int y~\mathrm{d}\Omega = \frac{\sigma_{\rm T}}{m_{\rm e} c^2} \frac{\smallint p\mathrm{d}V}{D^2_{\rm A}(z)},
\end{equation}
of $z < 1$ quiescent galaxies closely traces the simple self-similar scaling with halo mass,
\begin{equation}
  \label{eq:planck_scaling_corr}
Y~E^{-2/3}(z)~\left(\frac{D_\mathrm{A}(z)}{500~\mathrm{Mpc}}\right)^2 =(2.9 \pm 0.3) \times 10^{-8}
   \mathrm{arcmin}^2\left(\frac{M_\mathrm{h}}{10^{12} \mathrm{M}_\odot}\right)^{5/3}
\end{equation}
down to a halo mass of $\sim 3 \times 10^{13}$M$_\odot$. Broad agreement with this scaling relation
is found in \citet{gral14} when comparing the magnitude of their SZ signal to the expected halo mass
of radio-loud AGN hosts. The \emph{Planck} study gives this relation using the quantities $Y_{500}$
and $M_{500}$, defined with respect to a physical radius $R_{500}$ enclosing a mean density of
$500 \rho_{\rm c}(z)$ \citep[see also][]{lebr15}. In Equation~\ref{eq:planck_scaling_corr}, we have
accounted for a factor of $Y/Y_{500}\approx1.8$, where $Y$ corresponds to the total integrated SZ
signal. This factor is calculated using the same universal radial pressure profile
from~\citet{arna10} as is used in~\citet{plan13LBGSZ}. A factor of $M_{\rm h}/M_{500}=1.6$ is used,
consistent with the conversion from $M_{500}$ to $M_{200}$ from the concentration relation
of~\citet{duff08}. This yields a closer proxy of the halo virial mass referred to here as
$M_\mathrm{h}$. We assume the entirety of the signal we observe is enclosed within the ACT beams
such that $Y$, the total integrated signal is what our data constrain. This is a reasonable
assumption as $R_{200}$ for a $5 \times 10^{12}h^{-1}$M$_\odot$ halo corresponds to 0.4~arcmin
(within the ACT beam scales) at $z = 1.85$, the median redshift of our sample. Explicitly correcting
for the band dependent beam dilution effect for an SZ signal of this scale would require a model
dependent choice of SZ profile and result in corrections small relative to the measurement
uncertainties. Combining Equation~\ref{eq:sz_Y} and Equation~\ref{eq:planck_scaling_corr}, we find
\begin{equation}
  \label{eq:sz_scaling}
  \frac{\smallint p\mathrm{d}V}{10^{60} \mathrm{erg}} \approx 0.011~E^{2/3}(z) \left(\frac{M_\mathrm{h}}{10^{12}~ \mathrm{M}_\odot}\right)^{5/3}.
\end{equation}
While many of the assumptions made in deriving this relation may not hold for quasar hosts, these
arguments provide a rough estimate of the ${\smallint p\mathrm{d}V}$ signal one would expect from
purely gravitationally thermalized gas reservoirs hosted in these systems. For quasar halo masses in
the range $(1-5)\times 10^{12}h^{-1}$M$_\odot$ we find the expected level of this signal to be
${\smallint p\mathrm{d}V} = (0.4$--$5.8) \times 10^{59}$~erg at the median redshift of our
catalogue, $z=1.85$. Thus these purely gravitational arguments under predict our measured result of
${\smallint p\mathrm{d}V} = \pdvval \times 10^{60}$~erg making up at most $\sim$30 per cent of the
signal we observe, depending on the choice of halo mass. Given this amount of thermal energy the gas
in these haloes could either be gravitationally bound or unbound, again depending on the value taken
for the halo mass.

The validity of this calculation is dependent on the details of the high-mass tail of the halo mass
distribution of quasar hosts. Since in the case of gravitational heating, the SZ contribution of
each host halo is proportional to $M^{5/3}$, the effective mass that should be used in
Equation~\ref{eq:sz_scaling} is $\langle M^{5/3} \rangle^{3/5}$ where the average is taken over the
mass distribution of quasar host haloes. How much higher this effective mass is compared to the halo
masses we take from the literature depends on the behaviour of the high-mass tail of this
distribution. As an example of how this can affect the interpretation of these results, we
analytically approximate the average quasar host halo mass distribution of the $z\sim 1.4$ sample of
the \citet{rich12} clustering study. Using this approximation for the full mass distribution and
calculating $\langle M^{5/3} \rangle^{3/5}$, we find that the effective mass scale corresponds to an
expected signal from purely gravitational arguments of ${\smallint p\mathrm{d}V} \sim 4 \times 10^{60}$~erg.
Unfortunately, the high-mass end of the quasar halo mass distribution, upon which this calculation
relies, is not well constrained \citep{rich12, shen13}. \citet{shen13} find that data from clustering
studies can be fit equally well by models with very different parameterizations for the mass
dependence of the quasar halo occupation distribution, demonstrating the model dependent nature of
results for halo mass distribution in the literature (including the relative contribution of
correlated high-mass haloes to these studies). We therefore do not attempt to correct for this in
our primary discussion.

\subsection{Comparison with Other Studies}

\label{sec:comparison}

Recently \citet{ruan15} published a detection of quasar feedback in the \textit{Planck} data at the
level of $E_{\rm th} = 10^{62}$ erg, over an order of magnitude larger than one would expect from
the physical arguments presented here and what we measure. This level of SZ effect is characteristic
of the galaxy group mass scale and would leave a clear decrement at the level of several mJy around
each quasar in 150 GHz data from ACT and the SPT\@. Such a signal is challenged by both source
studies, such as our own, and power spectrum studies with ACT data \citep{dunk13, siev13}.
Furthermore, it appears that the feedback efficiency of 0.05 reported in \citet{ruan15} is not
computed as a fraction of a single accretion event's energy, as in our formalism, but rather as a
fraction of the total black hole rest mass energy. An energy of $10^{62}$~erg would correspond to
essentially 100 per cent efficient feedback. Another issue with this level of $E_{\rm th}$ is that
it exceeds the binding energy
($E_{\rm b} \approx 2.5 \times 10^{59}~{\rm erg}~(M_{\rm h}/10^{12}~{\rm M}_\odot)^{5/3}
E^{2/3}(z)$) of the host haloes.

\citet{chat10} correlated \emph{WMAP} 5-Year data at 40, 60, and 90~GHz with a large photometric
catalogue of quasars from the SDSS Data Release 3. They reported a 2.5$\sigma$ flux decrement at
40~GHz and estimated a mJy integrated SZ effect from a two parameter fit (SZ and dust) to their
three-band SED\@. While this amplitude is not dissimilar to that found here, \citet{chat10} in their
discussion noted the difficulty of establishing a firm conclusion about the nature of their
measurement given the limited frequency coverage, sensitivity and resolution of the millimetre data.
In particular, significant systematic effects were associated with the choice of foreground
treatment and data selection, likely resulting in observed systematics at low frequency where much
of their detection significance was derived \citep[cf. figs 7.12--15 of][]{chat09}.

\citet{cen15b} uses a halo catalogue from the \textit{Millenium} simulation with analytic
prescriptions for identifying quasar hosts (outlined in \citet{cen15a}) and constructing an
associated Compton-$y$ map. Their study demonstrates that contaminating projected SZ signal from two
halo correlations is important to account for in quasar stacking studies, particularly for large
beam sizes, $\sim 10$~arcmin. This effect is less important for higher resolution, $\sim 1$~arcmin
(similar to the ACT beam scale) studies but may need to be modelled in future work. Additionally
they find that such high-resolution studies may be promising in constraining models for the quasar
halo occupation distribution.

During the final stage of preparing this document, \citet{verd15} submitted a similar study of the
SDSS DR12 quasar catalogue using 70--857~GHz data from the \textit{Planck} satellite. Their measured
SZ amplitude is broadly consistent with that found in this work, however, there are some differences
in the quasar dust properties and interpretation of the SZ effect.

\section{Conclusions}
\label{sec:conclusions}
Using a stacking analysis of ACT and \emph{Herschel} data with the SDSS spectroscopic quasar
catalogue we reconstruct millimetre and FIR SEDs of quasars spanning the redshift range
$0.5 < z < 3.5$. We fit these data with a model for the stacked SED incorporating a dust component
that is allowed to evolve as a function of redshift as well as an SZ distortion. Our conclusions are
as follows:

\begin{itemize}
\item While the observed signal is dust dominated, 3$\sigma$--4$\sigma$ evidence is found for the
  thermal SZ effect associated with these systems for most dust models. In
  Appendix~\ref{apsec:alt_models}, we explore the effects of alternative dust parameterizations on
  this result.

\item This observed SZ signal is consistent with the scenario that this energy is being fuelled by
  quasar feedback in which up to $ \fval \tau_8^{-1}$ per cent of the quasar radiative energy is
  thermalized in the surrounding medium.

\item Fitting for the typical thermal energy associated with quasars, we find a best-fitting value
  of $E_{\rm th} = \pdvtotval \times 10^{60}$~erg. This exceeds by an order of magnitude what would
  be expected for a gravitationally heated circumgalactic medium for the
  $(1-5)\times 10^{12}h^{-1}$M$_\odot$ haloes quasars are though to occupy. However, if correct, the
  highest quasar halo mass estimates found in the literature could explain a large fraction of the
  observed SZ effect with purely gravitational heating.

\item Using the relation from \citet{syme08} and the rest-frame 160~\micron~luminosities from our
  SED models, we determine upper limits for the average star formation rates of quasar hosts of
  $\sim 60,~250$ and $ 400~{\rm M}_{\odot}/{\rm yr}$ for $z \sim $1, 2 and 3, respectively.
\end{itemize}

Forthcoming, deeper and wider ACT data will provide more precise estimates for the SEDs in the
millimetre but without a better understanding of the halo masses and dust properties of high
redshift quasar hosts, future studies will remain limited by associated systematics in the
interpretation of the improved constraints.

\section*{Acknowledgements}
We thank the anonymous referee for their useful comments and suggestions. This work was supported by
the US National Science Foundation through awards AST-0408698 and AST-0965625 for the ACT project,
as well as awards PHY-0855887 and PHY-1214379. Funding was also provided by Princeton University,
the University of Pennsylvania, and a Canada Foundation for Innovation (CFI) award to UBC\@. ACT
operates in the Parque Astron\'omico Atacama in northern Chile under the auspices of the Comisi\'on
Nacional de Investigaci\'on Cient\'ifica y Tecnol\'ogica de Chile (CONICYT). Computations were
performed on the GPC supercomputer at the SciNet HPC Consortium. SciNet is funded by the CFI under
the auspices of Compute Canada, the Government of Ontario, the Ontario Research Fund -- Research
Excellence; and the University of Toronto. We acknowledge the use of the Legacy Archive for
Microwave Background Data Analysis (LAMBDA), part of the High Energy Astrophysics Science Archive
Center (HEASARC). HEASARC/LAMBDA is a service of the Astrophysics Science Division at the NASA
Goddard Space Flight Center. This research made use of Astropy, a community-developed core {\sc python}
package for Astronomy \citep{astr13} and the affine invariant MCMC ensemble sampler implementation
provided by the {\sc emcee} {\sc python} package \citep{fore13}.

\bibliographystyle{mnras}
\bibliography{qsostack}

\appendix

\section{Optical Bolometric Luminosity Determination}
\label{apsec:lum_det}
\begin{figure}
\centering
\includegraphics[width=\columnwidth]{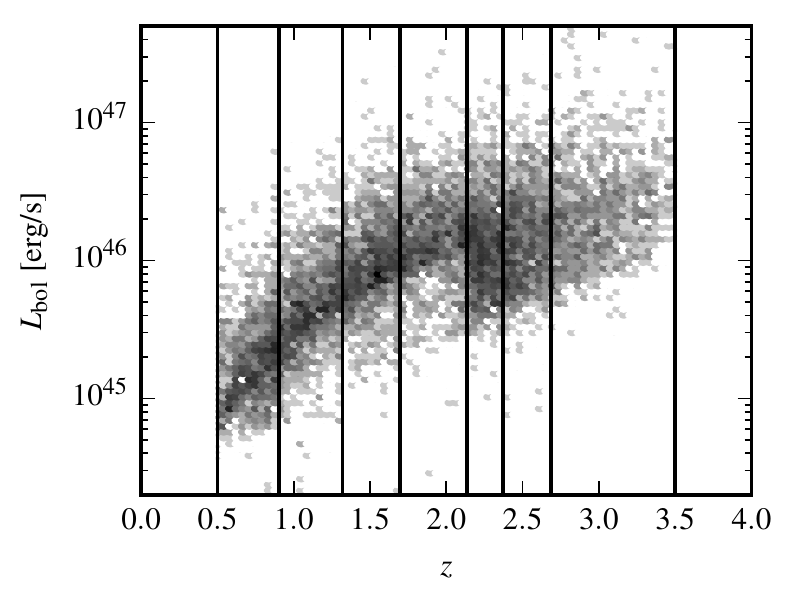}
\caption{The optical bolometric luminosity--redshift distribution of combined SDSS DR7 and DR10
  spectroscopic quasar sample falling within the ACT equatorial region
  ($-55\degr \leq \alpha \leq 58\degr$, $-1\fdg5 \leq \delta \leq 1\fdg5$) after the radio-loud cut
  has been applied. The density grey-scale used is logarithmic and the vertical black lines denote
  the boundaries of the redshift bins we use for the stacking analysis. The optical bolometric
  luminosity--redshift distribution of the subset of our sample which lies within the HerS region is
  found to be similar.
  \label{fig:lbol_z}}
\end{figure}

As we investigate models that assume the thermal energy we observe through the SZ effect is
dominated by converted quasar radiative energy (see the quasar feedback model described in
Section~\ref{sec:modelling}), we additionally construct estimates for the optical bolometric
luminosity, $L_{\rm bol}$, of each of the quasars in this sample. For this purpose we again make use
of SDSS $i$-band absolute magnitudes, $\mathcal{M}_I (z=2)$ which are converted to rest-frame
luminosities $L_{2500}$ at 2500~\AA~and then to $L_{\rm bol}$ by applying the bolometric correction
from \citet{rich06a}. These estimated bolometric luminosities are shown in Fig.~\ref{fig:lbol_z}.
Since the DR10 survey was designed to target quasars at $z>2$ its fainter magnitude limit
($i < 20.5$) yields an increase in lower luminosity quasars at high redshift \citep{pari14} observed
as the discontinuity at $z\sim2$ in Fig.~\ref{fig:lbol_z}.

The bolometric correction varies in a systematic way as a function of luminosity and colour of the
quasar but these variations among the different sub-populations do not exceed 15\% \citep{rich06a}.
There are, however, systematic concerns in constructing these estimates. In particular, double
counting direct ultraviolet emission by also including re-radiated dust emission when integrating
the SED used to derive these values can lead to overestimation of the correction. By restricting the
integral of the SED to the range between 1\micron\ and 2~keV, \citet{kraw13} obtain a bolometric
correction of 2.75 from the 2500\AA\ monochromatic luminosity compared to the value of 5 from
\citet{rich06a}, providing a lower bound on the bolometric correction. \citet{kraw13} further
explore a range of models and observations for the ultraviolet and X-ray spectra of quasars finding
bolometric corrections from 2500\AA\ to lie between 2.75 and 5. Using a different sample and
different methods, \citet{marc04} calculate bolometric corrections from $B$ band to be between 5 and
7, a value which weakly depends on quasar luminosity. Applying the \citet{vand01} power-law slope to
correct these to 2500\AA, we find bolometric corrections between 3.6 and 5.1. A similar procedure
yields a range of bolometric corrections between 3.6 and 7.3 for the \citet{elvi94} SED. While we
make use of the \citet{rich06a} bolometric correction we note that these values are still not known
to better than 40 per cent, which systematically affects the interpretation of our SZ constraints on
quasar feedback.

\section{Optical Luminosity Dependence}
\label{apsec:lum_dep}
The feedback interpretation of the SZ signal operates under the assumption that a fraction of the
bolometric luminosity of these sources is being converted to thermal energy which we observe through
the SZ effect. This therefore implies that the most bolometrically luminous sources should dominate
in their contribution to our observed SZ distortion. This luminosity dependence is explicitly
accounted for in the quasar feedback model but in this appendix we explore an alternative means of
measuring this dependence by evaluating the effect of luminosity cuts in the quasar catalogue on the
SZ amplitude extracted from fitting to the $\pdv$ model. Simply cutting the 8435 objects with
optical bolometric luminosity $L_{\rm bol} > 1 \times 10^{46}$~erg (determined as described in
Appendix~\ref{apsec:lum_det}) we find that the best-fitting $\pdv$ shifts from
$\pdvval \times 10^{60}$~erg to $(1.3 \pm 0.6) \times 10^{60}$~erg. However, a na\"ive cut such as
this may produce biased results as it significantly alters the sample's redshift distribution due to
the strong redshift dependence of the optical bolometric luminosities (See Fig.~\ref{fig:lbol_z}).
We therefore perform a further test by selecting a range of redshift, $2.1 < z < 3.0$, over which
the luminosity distribution remains qualitatively similar. The objects falling within this range are
separated into two bins of bolometric luminosity, each of which is fit with a model including
greybody dust and an SZ component parameterized through $\pdv$. We find that for the low luminosity
bin ($L_{\rm bol} < 10^{46.3}$~erg) the best-fitting SZ amplitude is
$\pdv = 2.2\pm1.1 \times 10^{60}$~erg, whereas for the $L_{\rm bol} > 10^{46.3}$~erg bin we find
$\pdv = 4.4\pm1.8 \times 10^{60}$~erg, again demonstrating a marginal shift to higher SZ amplitude
for higher luminosity objects.

\section{Alternative Dust Models}
\label{apsec:alt_models}
The significance of the SZ effect depends on the marginalization of the parameters of the assumed
dust model in the fits. In this appendix we explore some alternative dust models and report how the
choice of dust model affects the SZ detection and the goodness of fit to the data. A summary of
these fits to alternative models is provided in Table~\ref{tab:alt_model_summary}.

\begin{table}
  \caption{Summary of fitting results for fiducial models as well as models with alternative dust treatments.}
\begin{threeparttable}
\noindent
\addtolength{\tabcolsep}{-1pt}
\addtolength{\extrarowheight}{1pt}
\begin{tabularx}{\columnwidth}{lcccccc}
  \hline
  Model description&
                     DoF&
                          $\chi^2$&
                                    PTE&
                                         $\Delta{\rm DoF}_{\rm ref}$\tnote{$a$}&
                                                                                 $\Delta\chi^2_{\rm ref}$\tnote{$a$}\\\hline
  \textbf{Fiducial models} &&&&& \\\hline
  Greybody only & 33 & \nullchisq & 0.02 & 0 & 0 \\
  Quasar feedback & 32 & \fchisq & 0.61 & 1 & \fnulldeltachisq \\
  $\pdv$ & 32 & \pdvchisq & 0.28 & 1 & \pdvnulldeltachisq \\
  \hline\multicolumn{7}{l}{\textbf{Alternative quasar feedback models}} \\ \hline
  Fixed $\beta$ \tnote{$b$} &&&&& \\
  $\quad \quad \beta = 1.6$ & 33 & 41.8 & 0.14  & 0 & 9.2  \\
  $\quad \quad \beta = 1.8$ & 33 & 53.3 & 0.01 & 0 & -2.3 \\
  Optically thick & 31 & 26.7 & 0.69 & 2 & 24.3\\
  \hline\multicolumn{7}{l}{\textbf{Alternative dust only models}} \\ \hline
  Two temperature & 26 & 28.1 & 0.36 & 7 & 22.9\\
  Broken power law & 31 & 39.4 & 0.14 & 2 & 11.6 \\
  \hline
\end{tabularx}
\addtolength{\tabcolsep}{1pt}
\addtolength{\extrarowheight}{-1pt}
\begin{tablenotes}
  \item[$a$] These values correspond to the reduction in degrees of freedom and $\chi^2$ of
    each model with respect to the reference model which is the fit with no SZ effect and a single
    temperature greybody dust spectrum (the ``Greybody Only" model).
  \item[$b$] These models are found to still prefer non-zero values at the $\sim3\sigma$ level in
    their marginalized constraints for the SZ amplitude. The negative $\Delta\chi^2_{\rm ref}$ for
    the $\beta=1.8$ model indicates it is a poorer fit than the reference greybody only model.
\end{tablenotes}
\label{tab:alt_model_summary}
\end{threeparttable}
\end{table}

\subsection{Fixed $\beta$ models}
\label{apsec:fixed_beta}
In order to understand the effect of taking the dust emissivity index $\beta$ as a free parameter in
our fits, we attempt fits with the value of $\beta$ fixed. As can be seen in
Fig.~\ref{fig:param_contours}, there is degeneracy between the best-fitting value of $\beta$ and
the efficiency of quasar feedback, parameterized through $f$. A similar degeneracy is seen in the
$\pdv$ model where, when fitting, the amplitude of the SZ signal is inversely proportional to the
fit value for $\beta$. In Section~\ref{subsec:dust_emissivity} we noted that our fit value of
$\beta$ is on the low end of the theoretically allowed range of 1--2, further motivating a check
using models with values more commonly seen in the literature. We perform the same modelling of the
stacked fluxes as described in Section~\ref{sec:modelling} but instead of allowing $\beta$ as a free
parameter we hold its value fixed. We use the quasar feedback parameterization for the SZ effect
here. Motivated by the results of \citet{beel06} and \citet{hard13} respectively, we fit models with
$\beta$ fixed to values of 1.6 and 1.8. As expected, these models produce worse $\chi^2$ values than
models in which $\beta$ is a free parameter as shown in Table~\ref{tab:alt_model_summary}. The model
with $\beta=1.6$ provides an adequate fit to the data with a PTE of 0.14. The $\beta=1.8$ model
provides a considerably poorer fit with a PTE of 0.01. However, while detected at a lower amplitude
($f \approx 8.5~\tau_8^{-1}$ per cent) the marginal constraints on the SZ amplitude for both of
these fixed $\beta$ models prefer non-zero values at the $\sim3\sigma$ level.

\subsection{Optically Thick Model}
We evaluate the effect of the assumption of dust optical depth $<< 1$ implicit in our fiducial dust
SEDs by fitting the data with a model where we allow for an optically thick component. The
corresponding dust SED for this optically thick model is

\begin{align}
  S_{\rm dust}(\nu, z, L_{\rm ir}, \tau_0, \beta, T_{\rm d}) &= \frac{L_{\rm ir}}{4\pi D_{\rm L}^2(z)}
  \frac{\phi((1+z)\nu)}{\int \phi(\nu^\prime) \mathrm{d}\nu^\prime}\\
  \phi(\nu) &= (1 - \mathrm{e}^{-\tau_0 (\nu/\nu_{100\mu})^\beta}) B_\nu(T),\\
\end{align}
where, $\nu_{100\mu}$ is defined such that $\tau_0$ represents the dust optical depth at rest-frame
wavelength of 100~\micron. We fit the data with $\tau_0,~\beta$ and $T_{\rm d}$ as the globally fit
dust parameters and use the quasar feedback parameterization of the SZ effect. We find this model
provides a good fit to the data (as summarized in Table~\ref{tab:alt_model_summary}) with
$\chi^2 = 26.7$ for 31 degrees of freedom, corresponding to a PTE of 0.69. Constraints on the SZ
effect, parameterized through $f$ are found to be unchanged. The dust opacity is not well
constrained with $\log_{10}{\tau_0} = 0.2^{+2.2}_{-0.5}$ but we find the expected positive
correlation between dust optical depth and dust emissivity index $\beta$ due to the broadening of
the dust SED peak provided by the optically thick model.

The marginalized best-fitting emissivity index for this model is $\beta = 1.3 \pm 0.3$ and the dust
temperature, $T_{\rm d} = 52 \pm 11$~K, is constrained to values centred higher but still
consistent with those determined in the optically thin models.

\subsection{Two Temperature Model}
\label{apsec:two_t}
In this section, we examine whether the interpreted SZ distortion could instead be the result of an
additional lower temperature dust component contributing to the SED at frequencies below the primary
greybody peak. To test this, we attempt a model with no SZ component and a dust component that is
made up of the sum of two greybody SEDs. Such models were found to be highly degenerate in the
parameters describing the two dust components. To overcome some of this degeneracy we fix the
temperature of the primary (high temperature) dust component to 40~K, consistent with what we find
in the quasar feedback and $\pdv$ models. We then fit for a single value for the temperature of the
cold dust component across all redshift bins, restricting it to be $< 30$~K with a truncated uniform
prior. We fit the amplitudes of both components, parameterized through their infrared bolometric
luminosity, $L_{\rm ir}$ as defined in Section~\ref{sec:modelling}, on a bin by bin basis. We find
that the resulting best-fitting model provides a good fit to the data with a $\chi^2=28.1$ for 26
degrees of freedom, corresponding to the PTE of 0.35. The best-fitting value for the temperature of
the cold dust component corresponds to $20\pm3$~K and the bolometric luminosities for this component
range from $10^{10.7-11.4}$~L$_\odot$, increasing with redshift. This two temperature model cannot
be ruled out as an alternative explanation for the signal we observe. However (as shown in
Table~\ref{tab:alt_model_summary}) this reduction in $\chi^2$ of 22.9 with respect to the reference
greybody dust only model comes at the expense of seven additional parameters, whereas both SZ models
are significantly less complex, producing comparable $\chi^2$ reductions for a single additional
parameter. Applying the Bayesian information criterion (BIC), a test that explicitly penalizes
additional model complexity, we find a $\Delta$BIC improvement of $>$10. This corresponds to
``strong evidence" against the preference of this two temperature model over the quasar feedback
model based on this criterion.

\subsection{Broken power-law Model}
We additionally attempt fits using a dust model that has two emissivity indices, such that the dust
spectrum takes the form of a broken power law at low frequencies.

The modelled dust spectrum then becomes
\begin{align}
  S_{\rm dust}(\nu, z, L_{\rm ir}, \nu_{\rm b}, \beta_0, \beta_1, T_{\rm d}) = \frac{L_{\rm ir}}{4\pi D_{\rm L}^2(z)}
  \frac{\phi((1+z)\nu)}{\int \phi(\nu^\prime) \mathrm{d}\nu^\prime}\\
  \phi(\nu) = \begin{cases}(\nu/\nu_{\rm b})^{\beta_0} B_{\nu}(T_{\rm d})\ {\rm for}\ \nu \leq \nu_{\rm b} \\
    (\nu/\nu_{\rm b})^{\beta_1} B_{\nu}(T_{\rm d})\ {\rm for}\ \nu > \nu_{\rm b}\\
              \end{cases},
\end{align}
where $\nu_{\rm b}$ parameterizes the location of the spectral break with $\beta_0$ and $\beta_1$
parameterizing the emissivity index above and below this break respectively. We fit a model with
only a dust component of this form to determine whether the SZ distortion we measure could instead
be explained by such a spectrum. In this model, we fit single values for
$\nu_{\rm b},\ \beta_0,\ \beta_1$ and $T_{\rm d}$ while the $L_{\rm ir}$ values are fit
independently on a bin by bin basis. As shown in Table~\ref{tab:alt_model_summary}, the resulting
$\chi^2$ of 39.4 with 31 degrees of freedom (corresponding to a PTE of 0.14) indicates an adequate
fit to the data. However, the best-fitting constraints on the emissivity indices are found to be
$\beta_0 = 1.52^{+0.21}_{-0.25}$ and $\beta_1= 0.25^{+0.18}_{-0.31}$ with the rest-frame spectral
break at $\nu_{\rm b} = 1250^{+170}_{-210}$~GHz. This value for $\beta_1$ is well outside the
theoretically allowed region of $1 < \beta < 2$ as discussed in Section~\ref{subsec:dust_emissivity}
and therefore we reject this fit as less likely on physical grounds.

\bsp
\label{lastpage}
\end{document}